\documentclass[10pt,letterpaper]{article}
\usepackage[top=0.85in,left=2.75in,footskip=0.75in]{geometry}
\usepackage{xcolor}
\usepackage{bm}
\usepackage{changepage}

\usepackage[utf8x]{inputenc}

\usepackage{textcomp,marvosym}

\usepackage{fixltx2e}

\usepackage{amsmath,amssymb}

\usepackage{cite}

\usepackage{nameref,hyperref}

\usepackage[right]{lineno}

\usepackage{microtype}
\DisableLigatures[f]{encoding = *, family = * }

\raggedright
\usepackage[aboveskip=1pt,labelfont=bf,labelsep=period,justification=raggedright,singlelinecheck=off]{caption}

\makeatletter
\renewcommand{\@biblabel}[1]{\quad#1.}
\makeatother

\date{}

\usepackage{lastpage,fancyhdr,graphicx}
\usepackage{epstopdf}
\fancyheadoffset[L]{2.25in}
\fancyfootoffset[L]{2.25in}
\begin{document}
\vspace*{0.2in}

\begin{flushleft}
{\Large
\textbf\newline{Users Polarization on Facebook and Youtube} % Please use "title case" (capitalize all terms in the title except conjunctions, prepositions, and articles).
}
\newline
\\
Alessandro Bessi\textsuperscript{1,2},
Fabiana Zollo\textsuperscript{2},
Michela Del Vicario\textsuperscript{2},
Michelangelo Puliga\textsuperscript{2},
Antonio Scala\textsuperscript{2},
Guido Caldarelli\textsuperscript{2},
Brian Uzzi\textsuperscript{4\ddag},
Walter Quattrociocchi\textsuperscript{2,3*},

with the Lorem Ipsum Consortium\textsuperscript{\textpilcrow}
\\
\bigskip
\textbf{1} IUSS, Pavia, Italy
\\
\textbf{2} CSSLab, IMT Lucca, Italy
\\
\textbf{3} ISC, CNR, Rome, Italy
\\
\textbf{4} NICO, Northwestern University, Evanston, IL, USA
\\
\bigskip

% Insert additional author notes using the symbols described below. Insert symbol callouts after author names as necessary.
% 
% Remove or comment out the author notes below if they aren't used.
%
% Primary Equal Contribution Note
%\Yinyang These authors contributed equally to this work.

% Additional Equal Contribution Note
% Also use this double-dagger symbol for special authorship notes, such as senior authorship.
%\ddag These authors also contributed equally to this work.

%\textcurrency Current Address: Dept/Program/Center, Institution Name, City, State, Country % change symbol to "\textcurrency a" if more than one current address note
% \textcurrency b Insert second current address 
% \textcurrency c Insert third current address


% Group/Consortium Author Note
%\textpilcrow Membership list can be found in the Acknowledgments section.

% Use the asterisk to denote corresponding authorship and provide email address in note below.
* walterquattrociocchi@gmail.com

\end{flushleft}
\section*{Abstract}
Algorithms for content promotion accounting for users preferences might limit the exposure to unsolicited contents. In this work, we study how the same contents (videos) are consumed on different platforms -- i.e. Facebook and YouTube -- over a sample of $12M$ of users. 
Our findings show that the same content lead to the formation of echo chambers, irrespective of the online social network and thus of the algorithm for content promotion. 
Finally, we show that the users' commenting patterns are accurate early predictors for the formation of echo-chambers.

% Please keep the Author Summary between 150 and 200 words
% Use first person. PLOS ONE authors please skip this step. 
% Author Summary not valid for PLOS ONE submissions.   
%\section*{Author Summary}
%Lorem ipsum dolor sit amet, consectetur adipiscing elit. Curabitur eget porta erat. Morbi consectetur est vel gravida pretium. Suspendisse ut dui eu ante cursus gravida non sed sem. Nullam sapien tellus, commodo id velit id, eleifend volutpat quam. Phasellus mauris velit, dapibus finibus elementum vel, pulvinar non tellus. Nunc pellentesque pretium diam, quis maximus dolor faucibus id. Nunc convallis sodales ante, ut ullamcorper est egestas vitae. Nam sit amet enim ultrices, ultrices elit pulvinar, volutpat risus.

% Use "Eq" instead of "Equation" for equation citations.
\section*{Introduction}

The way people attempt to make sense of relevant issues changed with the shift from an era of mediated mass communication to one of disintermediated echo chambers \cite{Cacciatore2015,brown2007word, Richard2004, QuattrociocchiCL11,Quattrociocchi2014,Kumar2010}. On online social media, polarized communities emerge around diverse narratives. Some of these narratives reflect the extreme disagreement of public opinion on global and social issues. 
The emergence of polarization in online environments might reduce viewpoint heterogeneity, which has long been viewed as an important component of strong democratic societies \cite{dewey1927the,habermas1998between}.

Confirmation bias has been shown to play a pivotal role in the diffusion of rumors online \cite{del2016spreading}.
However, on online social media, different algorithms foster personalized contents according to user tastes -- i.e. they show users viewpoints that they already agree with, hence leading to the so called filter bubbles. 
Little is known about the factors affecting the algorithms' outcomes. Facebook promotes posts according to the \emph{News Feed} algorithm, that helps users to see more stories from friends they interact with the most, and the number of comments and likes a post receives and what kind of story it is -- e.g. photo, video, status update -- can also make a post more likely to appear \cite{newsfeed}. Conversely, YouTube promotes videos through \emph{Watch Time}, which prioritizes videos that lead to a longer overall viewing session over those that receive more clicks \cite{watchtime}. 
One hypothesis is that these algorithms might have a role in the emergence of echo chambers.
However,  not much is known about the role of cognitive factors in driving users to aggregate in echo chambers supporting their preferred narrative. Recent studies suggest confirmation bias as one of the driving forces of content selection, which eventually leads to the emergence of polarized communities \cite{bessi2014economy,bessi2014science,bessi2014viral,bessi2015trend,WEF16}.

In this work, we aim at characterizing the behavior of users dealing with the same contents, but different mechanisms of content promotion. By focusing on all YouTube videos posted by scientific and conspiracy-like Facebook pages, we want to understand whether different mechanisms regulating content promotion in Facebook and Youtube lead to the emergence of homogeneous echo chambers.

We choose to analyze such specific narratives for two main reasons: a) scientific news and conspiracy-like news are two very distinct and conficting narratives;
b) scientific pages share the main mission to diffuse scientific knowledge and rational thinking, while the alternative ones resort to unsubstantiated rumors.

Indeed, conspiracy-like pages disseminate myth narratives and controversial information, usually lacking supporting evidence and most often contradictory of the official news. Moreover,mthe spreading of misinformation on online social media has become a widespread phenomenon to an extent that the World Economic Forum listed massive digital misinformation as one of the main threats for the modern society \cite{howell2013digital,WEF16}.

In spite of different debunking strategies, unsubstantiated rumors -- e.g. those supporting anti-vaccines claims, climate change denials, and alternative medicine myths -- keep proliferating in polarized communities emerging on online enviroments \cite{bessi2014viral, del2016spreading}, leading to a climate of disengagement from mainstream society and recommended practices. A recent study \cite{zollo2015debunking} pointed out the inefficacy of debunking and the concrete risk of a backfire effect \cite{nyhan2010when, bessi2014social} from the usual and most commited consumers of conspiracy-like narratives.

We believe that additional insights about cognitive factors and behavioral patterns driving the emergence of polarized environments are crucial to understand and develop strategies to mitigate the spreading of online misinformation.

In this paper, using a quantitative analysis on a massive dataset ($12M$ of users), we compare consumption patterns of videos supporting scientific and conspiracy-like news on Facebook and Youtube. 

We extend our analysis by investigating the polarization dynamics -- i.e. how users become polarized comment after comment. On both platforms, we observe that some users interact only with a specific kind of content since the beginning, whereas others start their commenting activity by switching between contents supporting different narratives. The vast majority of the latter -- after the initial switching phase -- starts consuming mainly one type of information, becoming polarized towards one of the two conflicting narratives.

Finally, by means of a multinomial logistic model, we are able to predict with a good precision the probability of whether a user will become polarized towards a given narrative or she will continue to switch between information supporting competing narratives. The observed evolution of polarization is similar between Facebook and YouTube to an extent that the statistical learning model trained on Facebook is able to predict with a good precision the polarization of YouTube users, and vice versa.
Our findings show that conflicting narratives lead to the aggregation of users in different echo chambers, irrespective of the online social network and the algorithm of content promotion.

%The paper is structured as follows. First, we provide details about the data collection and some preliminary definitions. Then, we perform a comparative analysis to investigate the statistical signatures about distinct content consumption across Facebook and YouTube. Further, we show the emergence of polarized and homogeneous communities in both online social networks, and we provide a model that illustrates how users become polarized towards different narratives. 

\section*{Results}

We start our analysis by focusing on the statistical signatures of content consumption on Facebook and Youtube videos. The focus is on all videos posted by conspiracy-like and scientific pages on Facebook. We compare the consumption patterns of the same video on both Facebook and Youtube.
On Facebook a {\em like} stands for a positive feedback to the post; a {\em share} expresses the will to increase the visibility of a given information; and a {\em comment} is the way in which online collective debates take form around the topic promoted by posts. Similarly, on YouTube a {\em like} stands for a positive feedback to the video; and a {\em comment} is the way in which online collective debates grow around the topic promoted by videos.

\subsection*{Contents Consumption across Facebook and YouTube.}
\begin{figure*}[ht]
	\centering
	\includegraphics[width = \textwidth]{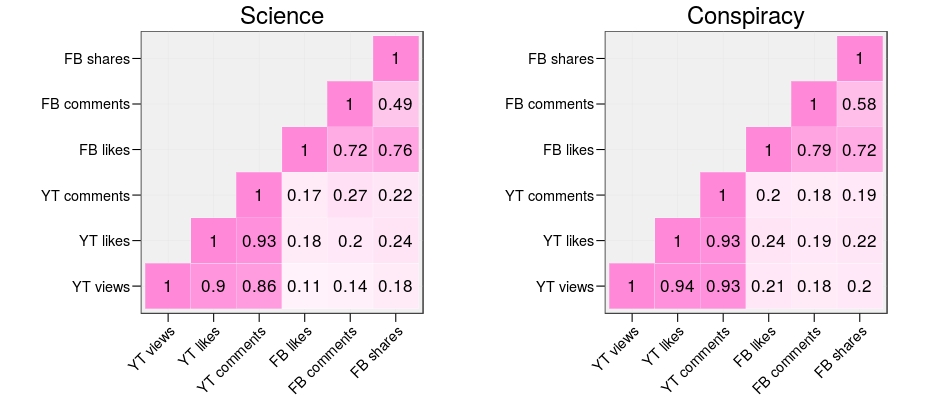}
	
	\caption{\textbf{Correlation Matrix.} Spearman's rank correlation coefficients between users' actions on Facebook posts and the related YouTube videos.}
	\label{fig:1}
\end{figure*}

Focusing on the consumptions patterns of YouTube videos posted on Facebook pages, we compute the Spearman's rank correlation coefficients between users' actions on Facebook posts and the related YouTube videos (see Figure \ref{fig:1}). 

By means of the Mantel test \cite{mantel1967} we find a statistically significant (simulated p-value $< 0.01$, based on $10^4$ Monte Carlo replicates), high, and positive ($r = 0.987$) correlation between the correlation matrices of Science and Conspiracy. In particular, we find positive and high correlations between users' actions on YouTube videos for both Science and Conspiracy, indicating a similar strong monotone increasing relationship between views, likes, and comments. Furthermore, we observe positive and mild correlations between users' actions on Facebook posts linking YouTube videos for both Science and Conspiracy, suggesting a monotone increasing relationship between likes, comments, and shares. Conversely, we find positive yet low correlations between users' actions across YouTube videos and the Facebook posts linking the videos for both Science and Conspiracy, implying that the success -- in terms of received attention -- of videos posted on YouTube does not ensure a comparable success on Facebook, and vice versa. Such results provide the first evidences towards a similar consumption behavior of users consuming conflicting narratives in different online social networks.

\begin{figure*}[ht]
	\centering
	\includegraphics[width = \textwidth]{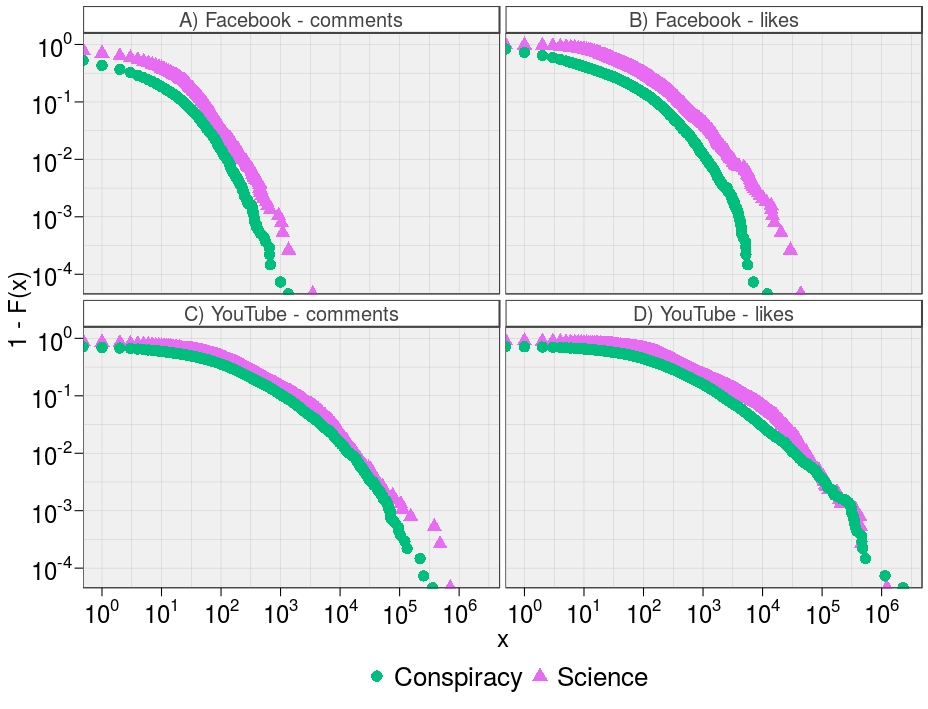}
	
	\caption{\textbf{Consumption Patterns of Videos on Facebook and YouTube.} The empirical CCDFs, $1 - F(x)$, show the consumption patterns of videos supporting conflicting narratives -- i.e. Science  and Conspiracy -- in terms of comments (A and C) and likes (B and D) on Facebook and YouTube.}
	\label{fig:2}
\end{figure*}

To further investigate users' consumption patterns, in Figure \ref{fig:2} we show the empirical Cumulative Complementary Distribution Functions (CCDFs) of the consumption patterns of videos supporting conflicting narratives -- i.e. Science  and Conspiracy -- in terms of comments and likes on Facebook and YouTube. The double-log scale plots highlight the power law behavior of each distribution. Top right panel shows the CCDFs of the number of likes received by Science ($x_{min} = 197$ and $\theta = 1.96$) and Conspiracy ($x_{min} = 81$ and $\theta = 1.91$) on Facebook. Top left panel shows the CCDFs of the number of comments received by Science ($x_{min} = 35$ and $\theta = 2.37$) and Conspiracy ($x_{min} = 22$ and $\theta = 2.23$) on Facebook. Bottom right panel shows the CCDFs of the number of likes received by Science ($x_{min} = 1,609$ and $\theta = 1.65$) and Conspiracy ($x_{min} = 1,175$ and $\theta = 1.75$) on YouTube. Bottom left panel shows the CCDFs of the number of comments received by Science ($x_{min} = 666$ and $\theta = 1.70$) and Conspiracy ($x_{min} = 629$ and $\theta = 1.77$) on YouTube. Our results indicate meaningful similarities in the consumption patterns of conflicting narratives on different online social networks.

\subsection*{Polarized and Homogeneous Communities.}
\begin{figure*}[ht]
	\centering
	\includegraphics[width = \textwidth]{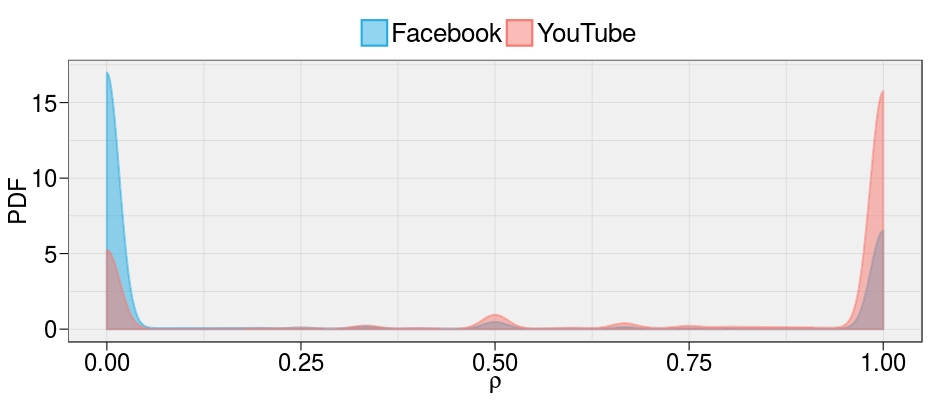}
	\caption{\textbf{Polarization on Facebook and YouTube.} The PDFs of the polarization $\rho$ show that the vast majority of users is polarized towards one of the two conflicting narratives -- i.e. Science and Conspiracy -- on both Facebook and YouTube.}
	\label{fig:4a}
\end{figure*}

We broaden our analysis by looking at how Facebook and Youtube users are polarized towards scientific or conspiracy-like contents. Figure \ref{fig:4a} shows the Probability Density Functions (PDFs) of about $12M$ users' polarization computed on Facebook and YouTube. We observe two bimodal distributions, indicating that most of the users are strongly polarized towards one of the two conflicting narratives in both online social networks. To quantify the degree of polarization we use the Bimodality Coefficient (BC), and we find that the BC is very high for both Facebook and YouTube. In particular, $BC_{FB} = 0.964$ and $BC_{YT} = 0.928$. Moreover, we observe that the percentage of polarized users (users with $\rho < 0.05$ and $\rho > 0.95$) is $93.6\%$ on Facebook and $87.8\%$ on YouTube; therefore, two well separated communities support competing narratives in both online social networks.

Such a result shows that conflicting narratives lead users to aggregate in well separated echo chambers, independently from the online social network and the specific algorithm of content promotion.

\begin{figure*}[ht]
	\centering
	\includegraphics[width = \textwidth]{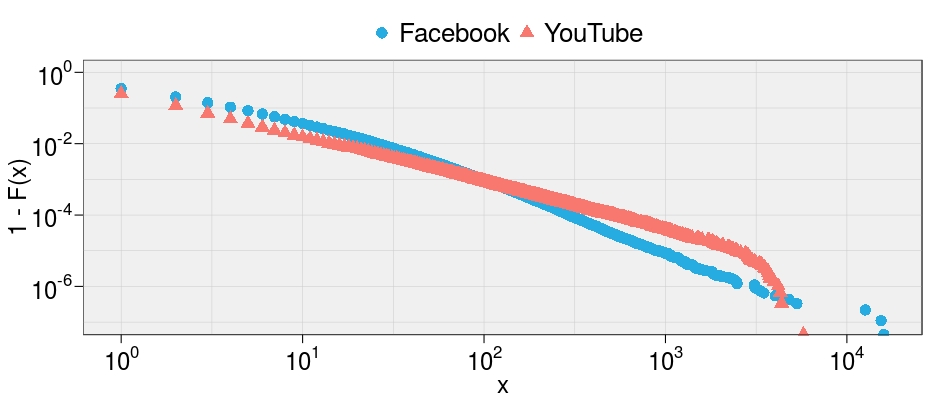}
	
	\caption{\textbf{Commenting Activity of Polarized Users.} The empirical CCDFs, $1 - F(x)$, of the number of comments left by polarized users on Facebook and YouTube.}
	\label{fig:4b}
\end{figure*}

To further characterize such a polarized environment, we analyze the consumption patterns of polarized users. Figure \ref{fig:4b} shows the empirical CCDFs of the number of comments left by all polarized users on Facebook and YouTube  ($x_{min}^{FB} = 8$, $\theta^{FB} = 2.13$ and $x_{min}^{YT} = 17$, $\theta^{YT} = 2.29$). We observe a very narrow difference ($\mathtt{HDI}90 = [-0.18,-0.13]$) between the tail behavior of the two distributions.

\begin{figure*}[ht]
	\centering
	\includegraphics[width = \textwidth]{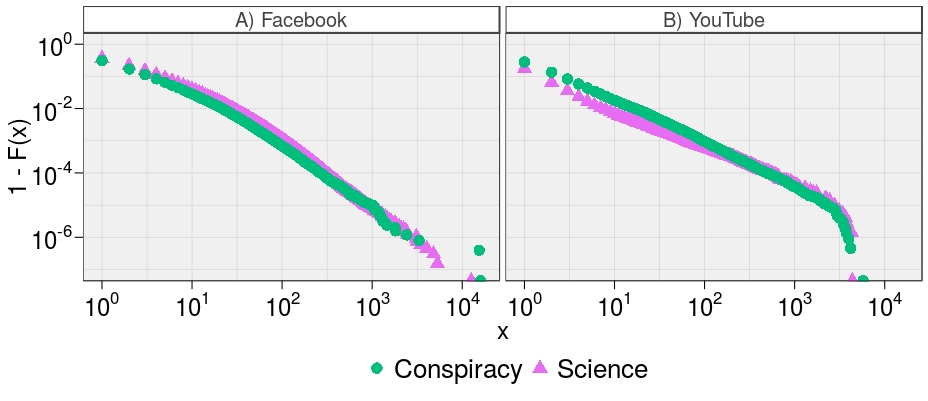}
	
	\caption{\textbf{Commenting Activity of Users Polarized towards Conflicting Narratives.} The empirical CCDFs, $1 - F(x)$, of the number of comments left by users polarized on scientific narratives and conspiracy theories on Facebook (A) and YouTube (B).}
	\label{fig:5}
\end{figure*}

Moreover, Figure \ref{fig:5} shows the empirical CCDFs of the number of comments left by users polarized on either Science or Conspiracy on both Facebook ($x_{min}^{Sci} = 5$, $\theta^{Sci} = 2.29$ and $x_{min}^{Con} = 4$, $\theta^{Con} = 2.31$, with $\mathtt{HDI}90 = [-0.018,-0.009]$) and YouTube ($x_{min}^{Sci} = 2$, $\theta^{Sci} = 2.86$ and $x_{min}^{Con} = 3$, $\theta^{Con} = 2.41$, with $\mathtt{HDI}90 = [0.44,0.46]$). Users supporting conflicting narratives behave similarly on Facebook, whereas on YouTube we observe a considerable difference between the scaling parameters of the distributions.

The aggregation of users around conflicting narratives lead to the emergence of echo chambers. Once inside such homogeneous and polarized communities, users supporting both narratives behave in a similar way, irrespective of the platform and the algorithm of content promotion.

\subsection*{Prediction of Users Polarization.}
We further extend our analysis by investigating the polarization dynamics -- i.e. how users' polarization evolves comment after comment.

We consider random samples of $400$ users who left at least $100$ comments, and we compute the mobility of a user across different contents along time. On both Facebook and YouTube, we observe that some users interact with a specific kind of content, whereas others start their commenting activity by switching between contents supporting different narratives. The vast majority of the latter -- after the initial switching phase -- starts consuming one type of information, becoming polarized towards one of the two conflicting narratives.

We exploit such a feature to derive a data-driven model to forecast users' polarizations. Indeed, by means of a multinomial logistic model, we are able to predict the probability of whether a user will become polarized towards a given narrative or she will continue to switch between information supporting competing narratives. 

In particular, we consider the users' polarization after $n$ comments, $\rho_{n}$ with $n = 1,\dots,100$, as a predictor to classify users in three different classes: Polarized in Science ($N = 400$), Not Polarized ($N = 400$), Polarized in Conspiracy ($N = 400$). 

\begin{figure*}[ht]
	\centering
	\includegraphics[width = \textwidth]{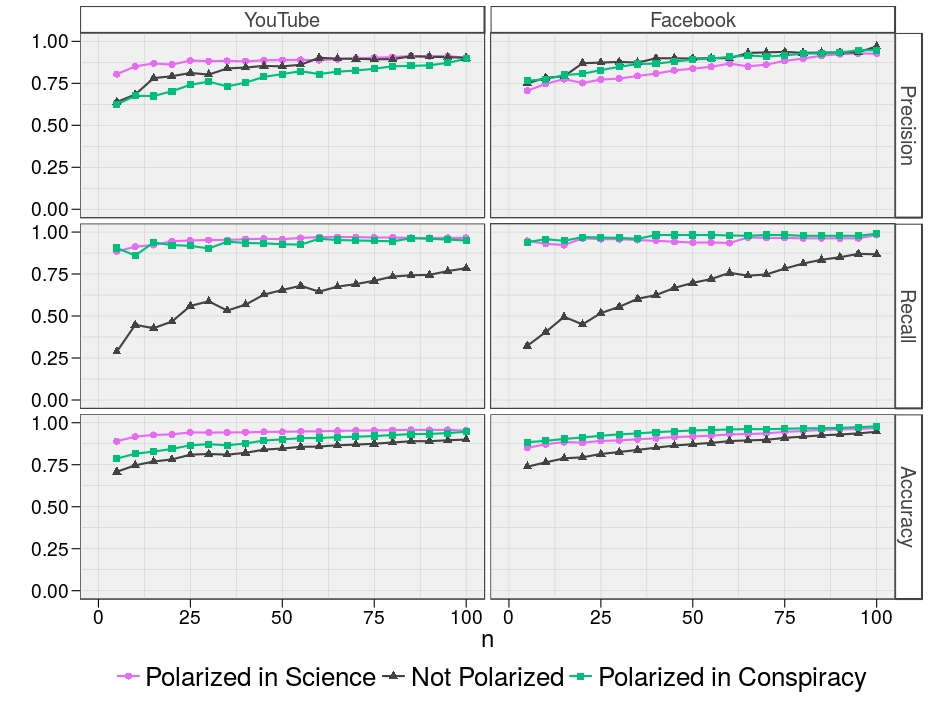}
	
	\caption{\textbf{Performance measures the classification task.} Precision, recall, and accuracy of the classification task for users Polarized in Conspiracy, Not Polarized, Polarized in Science on Facebook and YouTube as a function of $n$. On both online social networks, we find that the model's performance measures monotonically increase as a function of $n$. Focusing on the accuracy, significant results (greater than $0.70$) are obtained for low values of $n$.}
	\label{fig:8}
\end{figure*}

Figure \ref{fig:8} shows precision, recall, and accuracy of the classification tasks on Facebook and YouTube as a function of $n$. On both online social networks, we find that the model's performances monotonically increase as a function of $n$ for each class. Focusing on accuracy, significant results (greater than $0.70$) are obtained for low values of $n$. A suitable compromise between classification performances and required number of comments seems to be $n = 50$, which provides an accuracy greater than $0.80$ for each class on both YouTube and Facebook. 

To assess how the results generalize to independent datasets and to limit problems like overfitting, we split YouTube and Facebook users datasets in training sets ($N = 1000$) and test sets ($N = 200$), and we perform Monte Carlo cross validations with $10^3$ iterations. Results of Monte Carlo validations are shown in Table \ref{tab:1} and confirm the goodness of the model.

\begin{table}[ht]
	\centering
	\caption{\textbf{Monte Carlo Cross Validation.} Mean and standard deviation (obtained averaging results of $10^3$ iterations) of precision, recall, and accuracy of the classification task for users Polarized in Conspiracy, Not Polarized, Polarized in Science. }
	\label{tab:1}
	\begin{tabular}{ r  c | c | c || c | c | c}
		& \multicolumn{3}{c}{\textbf{YouTube}} & \multicolumn{3}{c}{\textbf{Facebook}} \\
		& Precision & Recall & Accuracy & Precision & Recall & Accuracy \\
		\hline
		Polarized in Conspiracy & $0.80\pm0.04$ &  $0.93\pm0.03$ & $0.90\pm0.02$ & $0.89\pm0.03$ & $0.98\pm0.02$ & $0.95\pm0.01$\\ 
		Not Polarized & $0.85\pm0.05$ &  $0.65\pm0.06$ & $0.85\pm0.02$ & $0.90\pm0.04$ & $0.70\pm0.05$ & $0.87\pm0.02$\\ 
		Polarized in Science & $0.89\pm0.04$ &  $0.96\pm0.02$ & $0.95\pm0.01$ & $0.84\pm0.04$ & $0.94\pm0.03$ & $0.92\pm0.02$\\
		\hline
	\end{tabular}
	
\end{table}

We conclude that the early mobility on commenting is an accurate predictor of the preferential attachment of users to a specific echo chamber.

Moreover, in Table \ref{tab:2}, we show that the evolution of the polarization on Facebook and YouTube is so alike that 
the same model (with $n = 50$), when trained with Facebook users ($N = 1200$) to classify YouTube users ($N = 1200$), leads to an accuracy in the classification task greater than $0.80$ for each class. Similarly, using YouTube users as training set to classify Facebook users leads to similar performances. 

Such results highlight a strong similarity in behavioral patterns of users interacting in different online social networks.

\begin{table}[ht]
	\centering
	\caption{\textbf{Performance measures of classification.} Precision, recall, and accuracy of the classification task for users Polarized in Conspiracy, Not Polarized, Polarized in Science when YouTube users are used as training set to classify Facebook users (top table), and when Facebook users are used as training set to classify YouTube users (bottom table).}
	\label{tab:2}
	\begin{tabular}{ r  c | c | c }
		\multicolumn{4}{c}{\textbf{Training YouTube -- Test Facebook}} \\
		& Precision & Recall & Accuracy  \\
		\hline
		Polarized in Conspiracy & $0.90$ &  $0.95$ & $0.95$ \\ 
		Not Polarized & $0.90$ &  $0.41$ & $0.79$\\ 
		Polarized in Science & $0.68$ &  $1.00$ & $0.84$ \\
		\hline
		\multicolumn{4}{c}{}\\
		\multicolumn{4}{c}{}\\ 
		\multicolumn{4}{c}{\textbf{Training Facebook -- Test YouTube}} \\   
		\hline
		Polarized in Conspiracy & $0.77$ &  $0.96$ & $0.89$ \\ 
		Not Polarized & $0.72$ &  $0.69$ & $0.81$\\ 
		Polarized in Science & $0.97$ &  $0.77$ & $0.91$ \\
		\hline
	\end{tabular}
	
\end{table}
\section*{Discussion}

Algorithms for content promotion are supposed to be the main determinants of the polarization effect arising out of online social media. Still, not much is known about the role of cognitive factors in driving users to aggregate in echo chambers supporting their favorite narrative. Recent studies suggest confirmation bias as one of the driving forces of content selection, which eventually leads to the emergence of polarized communities \cite{bessi2014economy,bessi2014science,bessi2014viral,bessi2015trend}.

Our findings show that conflicting narratives lead to the aggregation of users in homogeneous echo chambers, irrespective of the online social network and the algorithm of content promotion.

Indeed, in this work, we characterize the behavioral patterns of users dealing with the same contents, but different mechanisms of content promotion. In particular, we investigate whether different mechanisms regulating content promotion in Facebook and Youtube lead to the emergence of homogeneous echo chambers.

We study how users interact with two very distinct and conflicting narratives -- i.e. conspiracy-like and scientific news -- on Facebook and YouTube.  Using extensive quantitative analysis, we find the emergence of polarized and homogeneous communities supporting competing narratives that behave similarly on both online social networks. Moreover, we analyze the evolution of polarization, i.e. how users become polarized towards a narrative. Still, we observe strong similarities between behavioral patterns of users supporting conflicting narratives on different online social networks. 

Such a common behavior allows us to derive a statistical learning model to predict with a good precision whether a user will become polarized towards a certain narrative or she will continue to switch between contents supporting different narratives. Finally, we observe that the behavioral patterns are so similar in Facebook and YouTube that we are able to predict with a good precision the polarization of Facebook users by training the model with YouTube users, and vice versa. 

\section*{Methods}

\subsection*{Ethics Statement.}
The entire data collection process has been carried out exclusively through the Facebook Graph API \cite{fb_graph_api} and the YouTube Data API \cite{youtube_api}, which are both publicly available, and for the analysis we used only public available data (users with privacy restrictions are not included in the dataset). The pages from which we download data are public Facebook and YouTube entities. User content contributing to such entities is also public unless the user's privacy settings specify otherwise and in that case it is not available to us.

\subsection*{Data Collection.}
The Facebook dataset is composed of 413 US public pages divided to Conspiracy and Science news. The first category (Conspiracy) includes pages diffusing alternative information sources and myth narratives -- pages which disseminate controversial information, usually lacking supporting evidence and most often contradictory of the official news. The second category (Science) includes scientific institutions and scientific press having the main mission of diffusing scientific knowledge. Such a space of investigation is defined with the same approach as in \cite{zollo2015debunking}, with the support of different Facebook groups very active in monitoring the conspiracy narratives. For both the categories of pages we downloaded all the posts (and their respective users interactions) in a timespan of 5 years (Jan 2010 to Dec 2014). To our knowledge, the final dataset is the complete set of all scientific and conspiracy-like information sources active in the US Facebook scenario up to date. 

The YouTube dataset is composed of about $17K$ videos linked by Facebook posts supporting Science or Conspiracy news. Videos linked by posts in Science pages are considered as videos disseminating scientific knowledge, whereas videos linked by posts in Conspiracy pages are considered as videos diffusing controversial information and supporting myth and conspiracy-like theories. Such a categorization is validated by all the authors and Facebook groups very active in monitoring conspiracy narratives. The exact breakdown of the data is shown in Table \ref{tab:3}.

\begin{table}[ht]
	\centering
	\caption{\textbf{Breakdown of the dataset.}}
	\label{tab:3}
	\begin{tabular}{ r  c | c | c }
		& \multicolumn{3}{c}{\textbf{Facebook}}  \\
		& Science & Conspiracy & Total \\
		\hline
	 Posts & $4,388$ &  $16,689$ & $21,077$ \\ 
	 Likes & $925K$ &  $1M$ & $1.9M$ \\ 
	 Comments & $86K$ &  $127K$ & $213K$ \\
 	 Shares & $312K$ &  $493K$ & $805K$ \\
		\hline
		& & & \\
			& \multicolumn{3}{c}{\textbf{YouTube}}  \\
			& Science & Conspiracy & Total \\
			\hline
			Videos & $3,803$ &  $13,649$ & $17,452$ \\ 
			Likes & $13.5M$ &  $31M$ & $44.5M$ \\ 
			Comments & $5.6M$ &  $11.2M$ & $16.8M$ \\
			Views & $2.1M$ &  $6.33M$ & $8.41M$ \\
			\hline
	\end{tabular}
\end{table} 

\subsection*{Preliminaries and Definitions.}
\paragraph*{Polarization of Users.}
Polarization of users, $\rho_{u} \in [0,1]$, is defined as the fraction of comments that a user $u$ left on posts (videos) supporting conspiracy-like narratives on Facebook (YouTube). In mathematical terms, given $s_{u}$, the number of comments left on Science posts by user $u$, and $c_{u}$, the number of comments left on Conspiracy posts by user $u$, the polarization of $u$ is defined as
$$\rho_{u} = \frac{c_{u}}{s_{u} + c_{u}}.$$
We then consider users with $\rho_{u} > 0.95$ as users polarized towards Conspiracy, and users with $\rho_{u} < 0.05$ as users polarized towards Science.

\paragraph*{Bimodality Coefficient.}
The Bimodality Coefficient (BC) \cite{pfister2013} is defined as 

$$ BC = \frac{\mu_{3}^{2} + 1}{\mu_{4} + 3\frac{(n-1)^2}{(n-2)(n-3)}}, $$

with $\mu_{3}$ referring to the skewness of the distribution and $\mu_{4}$ referring to its excess kurtosis, with both moments being corrected for sample bias using the sample size $n$.

The BC of a given empirical distribution is then compared to a benchmark value of $BC_{crit} = 5/9 \approx 0.555$ that would be expected for a uniform distribution; higher values point towards bimodality, whereas lower values point toward unimodality.

\paragraph*{Multinomial Logistic Model.}
Multinomial logistic regression is a classification method that generalizes logistic regression to multi-class problems, i.e. with more than two possible discrete outcomes \cite{greene2011econometric}. Such a model is used to predict the probabilities of the different possible outcomes of a categorically distributed dependent variable, given a set of independent variables.
In the multinomial logistic model we assume that the log-odds of each response follow a linear model
$$ \eta_{ij} = \log\left(\frac{\pi_{ij}}{\pi_{iJ}}\right) = \alpha_{j} + \bm{x}^{T}_{i}\bm{\beta}_{j},$$

where $\alpha_{j}$ is a constant and $\bm{\beta}_{j}$ is a vector of regression coefficients, for $j = 1,2,\dots,J-1$. Such a model is analogous to a logistic regression model, except that the probability distribution of the response is multinomial instead of binomial, and we have $J-1$ equations instead of one. The $J-1$ multinomial logistic equations contrast each of categories $j = 1,2,\dots,J-1$ with the baseline category $J$. If $J = 2$ the multinomial logistic model reduces to the simple logistic regression model.

The multinomial logistic model may also be written in terms of the original probabilities $\pi_{ij}$ rather than the log-odds. Indeed, assuming that $\eta_{iJ} = 0$, we can write
$$\pi_{ij} = \frac{\exp(\eta_{ij})}{\sum_{k=1}^{J}\exp(\eta_{ik})}.$$

\paragraph{Classification Performance Measures.}
To assess the goodness of our model we use three different measures of classification performance: precision, recall, and accuracy. For each class $i$, we compute the number of true positive cases $TP_{i}$, true negative cases $TN_{i}$, false positive cases $FP_{i}$, and false negative cases $FN_{i}$. Then, for each class $i$ the precision of the classification is defined as
$$ precision_{i} = \frac{TP_{i}}{TP_{i} + FP_{i}}, $$

the recall is defined as
$$ recall_{i} = \frac{TP_{i}}{TP_{i} + FN_{i}}, $$

and the accuracy is defined as
$$ accuracy_{i} = \frac{TP_{i} + TN_{i}}{TP_{i} + TN_{i} + FP_{i} + FN_{i}}. $$

\paragraph{Power law distributions.}
Scaling exponents of power law distributions are estimated via maximum likelihood (ML) as shown in \cite{clauset2009}. To provide a full probabilistic assessment about whether two distributions are similar, we estimate the posterior distribution of the difference between the scaling exponents through an Empirical Bayes method.

Suppose we have two samples of observations, $A$ and $B$, following power law distributions.
For the sample $A$, we use the ML estimate of the scaling parameter, $\hat{\theta}_{A}^{ML}$, as location hyper-parameter of a Normal distribution with scale hyper-parameter $\hat{\sigma}_{A}^{ML}$. Such a Normal distribution represents the prior distribution, $p(\theta_{A}) \sim \mathcal{N}(\hat{\theta}_{A}^{ML}, \hat{\sigma}_{A}^{ML})$, of the scaling exponent $\theta_{A}$. Then, according to the Bayesian paradigm, the prior distribution, $p(\theta_{A})$, is updated into a posterior distribution, $p(\theta_{A}|x_{A})$:
$$ p(\theta_{A}|x_{A}) = \frac{p(x_{A}|\theta_{A})p(\theta_{A})}{p(x_{A})},$$

where $p(x_{A}|\theta_{A})$ is the likelihood. The posterior distribution is obtained via Metropolis-Hastings algorithm, i.e. a Markov Chain Monte Carlo (MCMC) method used to obtain a sequence of random samples from a probability distribution for which direct sampling is difficult \cite{gelman2003bayesian,andrieu2003introduction,hartig2011statistical}. To obtain reliable posterior distributions, we run $50,000$ iterations ($5,000$ burned), which proved to ensure the convergence of the MCMC algorithm.

The posterior distribution of $\theta_{B}$ can be computed following the same steps.
Once both posterior distributions, $p(\theta_{A}|x_{A})$ and $p(\theta_{B}|x_{B})$, are derived, we compute the distribution of the difference between the scaling exponents by subtracting the posteriors, i.e. 
$$p(\theta_{A} - \theta_{B}|x_{A},x_{B}) = p(\theta_{A}|x_{A}) - p(\theta_{B}|x_{B}).$$

Then, by observing the $90\%$ High Density Interval ($\mathtt{HDI}90$) of $p(\theta_{A} - \theta_{B})$, we can draw a full probabilistic assessment of the similarity between the two distributions.

\bibliography{biblio_final}

\section*{Acknowledgements}
Funding for this work was provided by EU FET project MULTIPLEX nr. 317532, SIMPOL nr. 610704, DOLFINS nr. 640772, SOBIGDATA nr. 654024. The funders had no role in study design, data collection and analysis, decision to publish, or preparation of the manuscript. We want to thank Geoff Hall and “Skepti Forum” for providing fundamental support in defining the atlas of conspiracy news sources in the US Facebook.

\section*{Author contributions statement}

A.B. and W.Q. conceived the experiment. A.B. and M.P. equally contributed to the simulations and the data analysis. All the authors equally contributed to the writing of the paper.

\section*{Additional information}

\textbf{Competing financial interests} The authors declare no competing financial interests.

\nolinenumbers

% Either type in your references using
% \begin{thebibliography}{}
% Text
% \end{thebibliography}
%
% or
%
% Compile your BiBTeX database using our plos2015.bst
% style file and paste the contents of your .bbl file
% here.
% 

\end{document}